\documentstyle [seceq,psfig,preprint] {ptptex}
\def\vecx{\mbox {\boldmath $x$}}

\preprintnumber[3cm]{SMUHEP/00-19}

\title{ Extended Hamiltonian Formalism of the\\ 
Pure Space-Like Axial Gauge Schwinger Model }

\author{ Yuji {\sc Nakawaki} and Gary {\sc McCartor}$^*$
}

\inst{Division of Physics and Mathematics, Faculty of Engineering \\
 Setsunan University, Osaka 572-8508, Japan
\\
$^*$Department of Physics, Southern Methodist University  \\
Dallas TX 75275,  USA}

\recdate{
}

\abst{  We demonstrate that  pure space-like axial gauge quantizations of 
gauge fields can be constructed in ways which are free from infrared divergences. To do so we must extend the Hamiltonian formalism to include residual gauge fields. We 
construct an operator solution and an extended Hamiltonian of the pure 
space-like axial gauge Schwinger model. We begin by constructing an axial 
gauge formulation in auxiliary 
coordinates: $x^{\mu}=(x^+,x^-)$, where $x^+=x^0\sin{\theta}+x^1\cos{\theta},
\; x^-=x^0\cos{\theta}-x^1\sin{\theta}$ and we take $A_-=A^0\cos{\theta}+A^1
\sin{\theta}=0$ as the gauge fixing condition. In the region $0{\le}
{\theta}<\frac{\pi}{4}$, we can take $x^-$ as the evolution  parameter and 
construct a traditional canonical formulation of the temporal gauge Schwinger model 
in which residual gauge fields dependent only on $x^+$ are static 
canonical variables. Then we extrapolate the temporal gauge operator solution 
into the axial region, $\frac{\pi}{4}<{\theta}<\frac{\pi}{2}$, where $x^+$ is 
taken as the evolution parameter. In the axial region we find that we have to 
change representations of the residual gauge fields from one realizing the PV 
prescription to one realizing the ML prescription in order for the infrared 
divergences resulting from $({\partial}_-)^{-1}$  to be canceled by 
corresponding ones resulting from the inverse of the hyperbolic Laplace operator. We 
overcome the difficulty of constructing the Hamiltonian for the residual gauge 
fields by employing McCartor and Robertoson's method, which gives us a term integrated over $x^-=$ constant. Finally, by taking the limit 
${\theta}\to\frac{\pi}{2}-0$ we obtain an operator solution and the Hamiltonian of the 
axial gauge (Coulomb gauge )Schwinger model in ordinary coordinates. That solution includes auxiliary fields and the representation space is of indefinite metric, providing further evidence that ``physical'' gauges are no more physical than ``unphysical'' gauges.}   

\begin{document}

\maketitle

\section{Introduction}
Axial gauges, $n^{\mu}A_{\mu}=0$, specified by a constant vector $n^{\mu}$, 
have been used recently in spite of their lack of manifest Lorentz 
covariance. One reason is that the Faddeev-Popov ghosts decouple from 
the theory in the axial gauge formulations.$^{1)}$ The case of $n^2=0$, the 
light-cone gauge, has been extensively used in light-front field theory (LFFT) 
in attempts to find nonperturbative solutions of QCD. In LFFT one usually 
makes the change of variables, $x^+_l=\frac{x^0+x^3}{\sqrt{2}},\;x^-_l
=\frac{x^0-x^3}{\sqrt{2}},^{2)}$, and specifies quantization conditions on 
the hyperplane $x^+_l = 0$.  Advantages are that the vacuum state contains 
only of particles with nonnegative longitudinal momentum and that there exist 
relativistic bound-state equations of the Schr\"odinger-type. (For a good 
overview of LFFT see Ref. 3).)

It has been known for some time that axial gauge formulations are not ghost 
free, contrary to what was originally expected and is still sometimes claimed. 
It was first pointed out by Nakanishi$^{4)}$ that there exists an intrinsic 
difficulty in the axial gauge formulations so that an indefinite metric is 
indispensable even in QED. It was also noticed that in order to bring 
perturbative calculations done in the light-cone gauge into agreement with 
calculations done in covariant gauges, spurious singularities of the free 
gauge field propagator have to be regularized not as principal values (PV), 
but according to the Mandelstam-Leibbrandt (ML) prescription$^{5)}$. Shortly 
afterwards, Bassetto et al.$^{6)}$ obtained the ML form of the propagator in a 
canonical formalism. They quantized at equal time in light-cone gauge, which 
required the introduction of a Lagrange multiplier field and its conjugate.  
These may be viewed  as residual gauge degrees of freedom. Furthermore, Morara 
and Soldati$^{7)}$ found just recently that the same is true in the light-cone 
temporal gauge formulation in which $x_l^+$ is again taken as the evolution 
parameter but $ \frac{A_0+A_3}{\sqrt{2}}=0$ is taken as the gauge fixing 
condition. McCartor and Robertson$^{8)}$ showed that in the light-cone axial 
gauge formulation, where $\frac{A_0-A_3}{\sqrt{2}}=0$ is taken as the gauge 
fixing condition, the translational generator $P_+$ consists of physical 
degrees of freedom integrated over the hyperplane $x_l^+=$ constant and 
residual gauge degrees of freedom integrated over the hyperplane $x_l^-=$ 
constant.

In McCartor and Robertson's work the residual degrees of freedom were viewed 
as integration constants necessary to completely specify the solution to the 
constraint equations which relate constrained degrees of freedom to the 
independent degrees of freedom.  That is the attitude we will adopt in the 
present paper.  Such equations often occur in quantum field theory and little 
attention has been given to the question of what boundary conditions are 
appropriate to completely specify their solution.  In the present paper we 
will attend to that question.  An example of the type of equation we have in 
mind is the equation for $A_+$ in the light-cone gauge
\begin{equation}
\partial_-^2A_- +\partial_-\partial_iA^i = J_-.
\end{equation}
This equation admits a solution of the form
$$
  C(x^+,\vecx^\perp) + x^- B(x^+,\vecx^\perp) + F(x),
$$
where $F(x)$ is any solution to the equation.  McCartor and Robertson showed 
that if $F(x)$ is taken to be the usual solution, which results from the 
replacement $\partial_\mu \rightarrow i k_\mu$ in the mode expansion, then 
neither $C(x^+,\vecx^\perp)$ or $B(x^+,\vecx^\perp)$ are zero in the case of 
QED.  On the other hand, in an an earlier work$^{9)}$ , we showed that for the 
Schwinger model in the light-cone gauge, $B(x^+,\vecx^\perp)$ is zero while 
$C(x^+,\vecx^\perp)$ is not.  It will be seen below that it is this last case 
that is peculiar: for all the solutions we shall give, both 
$B(x^+,\vecx^\perp)$ and $C(x^+,\vecx^\perp)$ are nonzero except for the 
special case of the light-cone gauge where $B(x^+,\vecx^\perp)$ happens to 
vanish.

Because the axial gauges can be viewed as continuous deformations of the 
light-cone gauge, extensions of the ML prescription outside the light-cone 
gauge formulation have also been studied. Lazzizzera$^{10)}$, and Landshoff 
and Nieuwenhuizen$^{11)}$, constructed canonical formulations for the non-pure 
space-like case ($n^0{\ne}0,n^2<0$) in ordinary coordinates. However, in spite 
of many attempts, no one has succeeded in constructing a consistent, pure 
space-like axial gauge ($n^0=0,\;n^2<0$) formulation. One problem is that when
the residual gauge fields are introduced as integration constants that occur 
when one integrates the axial gauge constraints, one cannot obtain their 
quantization conditions from the Dirac quantization procedure$^{12)}$. Another 
problem is that the residual gauge fields are independent of $x^3$ in the 
gauge $A_3=0$ so that the  Hamiltonian cannot be obtained by integrating 
densities made of those residual gauge fields over the three dimensional 
hyperplane $x^0=$ constant. 

This motivates us to study the problem of introducing the residual gauge 
fields into pure space-like axial gauge formulations. To construct 
the Hamiltonians and to determine quantization conditions, we construct the 
pure space-like gauge as a continuous deformation of the light-cone gauge. In 
a previous, preliminary work$^{13)}$ we have constructed an axial gauge 
formulation of noninteracting abelian gauge fields in the auxiliary 
coordinates $x^{\mu}=(x^+,x^-,x^1,x^2)$, where
\begin{equation}
 x^+=x^0{\rm sin}{\theta}+x^3{\rm cos}{\theta}, \quad
x^-=x^0{\rm cos}{\theta}-x^3{\rm sin}{\theta}  \label{eq:(1.1)}
\end{equation}
and $ A_-=A^0{\rm cos}{\theta}+A^3{\rm sin}{\theta}=0$
is taken as the gauge fixing condition. The same framework was used previously 
by others to analyze two-dimensional models.$^{14)}$ In the region $0{\le}
{\theta}<\frac{\pi}{4}$ we take $x^-$ as the evolution parameter and 
construct the canonical temporal gauge formulation; in that case the residual 
gauge fields are static canonical variables. By continuation, we can 
obtain operators satisfying the field equations in the axial region 
$\frac{\pi}{4}<{\theta}<\frac{\pi}{2}$, where $x^+$ is taken as the evolution 
parameter. In the axial region we cannot use the traditional way 
of constructing the Hamitonian for the residual gauge fields; we obtain it by 
integrating the divergence equation of the energy-momentum tensor over a 
suitable closed surface. Because the residual gauge fields do not depend on 
$x^-$, their boundary surface ($x^-\to{\pm}\infty)$ contributions have to be 
kept. As a consequence, we obtain the Hamiltonian which includes a part from 
integrating a density involving the residual gauge fields over $x^-=$ 
constant. Then, by taking the limit ${\theta}\to\frac{\pi}{2}-0$ we obtain the 
Hamiltonian in the axial gauge $A_3=0$ formulation in ordinary coordinates.  

In this paper we proceed to interacting gauge fields. As a first step in 
finding ways to introduce the residual gauge fields into interacting axial 
gauge theories, we consider the exactly solvable Schwinger model. Previous 
attempts to construct the axial gauge ($A_1=0$) operator solution to the 
Schwinger model have used a representation space isomorphic to (copies of) 
that of a positive metric free massive scalar field $\tilde{\Sigma}$. In other 
words, they have assumed that ($A_1=0$) is a physical gauge in that the entire 
representation space is physical. All such attempts encounter severe infrared 
difficulties.$^{15)}$. We expect that the difficulties can be overcome by 
introducing residual gauge fields depending only on $x^0$. 
To study that possibility, we consider constructing an operator solution of 
the pure space-like axial gauge Schwinger model as a continuation from the 
light-cone gauge. We begin by constructing an operator solution in the region 
$0{\le}{\theta}<\frac{\pi}{4}$. This is  straightforward since we can use the 
axial anomaly and the temporal gauge canonical 
quantization conditions as guiding principles in constructing the operator 
solution. We continue the operator solution into the axial region 
$\frac{\pi}{4}<{\theta}<\frac{\pi}{2}$. In the axial region the operator 
$({\partial}_-)^{-1}$ becomes singular and gives rise to infrared divergences. 
Furthermore, in the axial region the Laplace operator, $m^2-n_-{\partial}_+
^{\;\;2}$ (where $m^2=\frac{e^2}{\pi}$ and $n_-=\cos 2{\theta}$), becomes 
hyperbolic so that divergences result from its inversion. We find that the 
former divergences are canceled by the latter ones if we choose the 
representation of the residual gauge fields in such a way as to realize the ML 
prescription. As a consequence we recover well-defined fermion field 
operators. Bilinear products of these give rise to bosonized expressions for 
the vector current and the energy-momentum tensor through the equal $x^+$-time 
point-splitting procedure. We apply McCartor and Robertson's method$^{8)}$ and 
obtain the Hamiltonian for the residual gauge fields. It consists of a part 
from the physical operators integrated over $x^+=$ constant and a part from 
the residual gauge operators  integrated over $x^-=$ constant. This is the way 
we extend the traditional Hamiltonian formulation to obtain the Hamiltonian 
for the residual gauge fields. Now it is straightforward to obtain an operator 
solution and the Hamiltonian in the gauge $A_1=0$ (here, the Coulomb gauge) in 
the ordinary coordinates. All one has to do is to take the limit 
${\theta}\to\frac{\pi}{2}-0$. While that limit exists, we cannot simply set 
$\theta$ equal to $\frac{\pi}{2}$.  For that reason one might say that we have 
not really quantized the Schwinger model at equal time in the Coulomb gauge. 
On the other hand one might regard $\theta$ as a regulating parameter. In that 
view we see that, as was the case in the light-cone gauge$^{9,16)}$ , proper 
regulation of the theory necessarily involves information off the initial 
value surface. In any event, the solution contains the auxiliary fields and 
the representation space is of indefinite metric. This provides further 
indication that ``physical'' gauges are no more physical than ``unphysical'' 
gauges in that the representation space contains unphysical states and is of 
indefinite metric in all cases.

The paper is organized as follows: In ${\S}~2$, we use the axial anomaly and 
the temporal gauge quantization conditions as guiding principles to construct 
the temporal gauge operator solution of the Schwinger model in the auxiliary 
coordinates. We also point out that if we choose the representation of the 
residual gauge fields so as to realize the PV prescription, we can construct 
fermion field operators in such a way that vacuum expectation values of them 
vanish. In ${\S}~3$, we continue the solution into the axial 
region and show in detail that the infrared divergences inherent to the 
axial gauge quantizations are canceled if  we choose the 
representation of the residual gauge fields which realizes the ML 
prescription. We also discuss properties of the axial gauge solution. 
Section 4 is devoted to concluding remarks.

We use the following conventions: 
$$g_{--}=\cos 2{\theta}, \quad  g_{-+}=g_{+-}=\sin 2{\theta}, \quad 
g_{++}=-\cos 2{\theta},  $$
$$g^{--}=\cos 2{\theta}, \quad g^{-+}=g^{+-}=\sin 2{\theta},  \quad 
g^{++}=-\cos 2{\theta}, $$
$$n^{\mu}=(n^+,n^-)=(0,1), \quad 
n_{\mu}=(n_+,n_-)=(\sin 2{\theta},\cos 2{\theta}),
$$
$$ {\gamma}^0={\sigma}_1, \;{\gamma}^1=i{\sigma}_2, \; {\gamma}^5=-{\sigma}_3, 
$$
$$
{\gamma}^+={\gamma}^0\sin {\theta}+{\gamma}^1\cos {\theta}=
\left(
\begin{array}{cc} 
 0 & \sqrt{1+n_+}  \\
-{\varepsilon}(n_-)\sqrt{1-n_+} & 0 
\end{array}\right),$$
$${\gamma}^-={\gamma}^0\cos {\theta}-{\gamma}^1\sin {\theta}=
\left(
\begin{array}{cc}
0 & {\varepsilon}(n_-)\sqrt{1-n_+} \\
\sqrt{1+n_+} & 0
\end{array}\right).$$

\section{Temporal gauge formulation in the auxiliary coordinates}

In this section we confine ourselves to the region $0{\le}{\theta}<\frac{\pi}
{4}$ and choose $x^-$ as the evolution parameter. The temporal gauge Schwinger 
model in the auxiliary coordinates is defined by the Lagrangian 
\begin{equation}
L=-\frac{1}{4}F_{{\mu}{\nu}}F^{{\mu}{\nu}}-B(n{\cdot}A) 
+i\bar{\psi}{\gamma}^{\mu}({\partial}_{\mu}+ieA_{\mu}){\psi}  
\label{eq:(2.1)}
\end{equation} 
where $B$ is the Lagrange multiplier field, that is, the Nakanishi-Lautrup 
field in noncovariant formulations.$^{17)}$ From the Lagrangian we derive the 
field equations
\begin{equation}
{\partial}_{\mu}F^{{\mu}{\nu}}=n^{\nu}B +J^{\nu}, \quad 
J^{\nu}=e\bar{\psi}{\gamma}^{\nu}{\psi} \label{eq:(2.2)}
\end{equation}
\begin{equation}
i{\gamma}^{\mu}({\partial}_{\mu}+ieA_{\mu}){\psi}=0, \label{eq:(2.3)}
\end{equation}
and the gauge fixing condition
\begin{equation}
A_-=0. \label{eq:(2.4)}
\end{equation}
The field equation of $B$,
\begin{equation}
{\partial}_-B=0, \label{eq:(2.5)}
\end{equation}
is obtained by operating on (\ref{eq:(2.2)}) with ${\partial}_{\nu}$.
 Canonically conjugate momenta are defined by 
\begin{equation}
{\pi}^-=\frac{{\delta}L}{{\delta}{\partial}_-A_-}=0,\;
{\pi}^+=\frac{{\delta}L}{{\delta}{\partial}_-A_+}=F_{-+},\;
{\pi}_B=\frac{{\delta}L}{{\delta}{\partial}_+B}=0,\;
{\pi}_{\psi}=\frac{{\delta}L}{{\delta}{\partial}_-{\psi}}=i\bar{\psi}{\gamma}^-
.   \label{eq:(2.6)}
\end{equation}
Thus we impose the following equal $x^-$-time canonical quantization 
conditions:
\begin{equation}
[A_+(x),A_+(y)]=[{\pi}^+(x),{\pi}^+(y)]=0, \quad [A_+(x),{\pi}^+(y)]=
i{\delta}(x^+-y^+),  \label{eq:(2.7)}
\end{equation}
\begin{equation}
[A_+(x),{\psi}(y)]=[{\pi}^+(x),{\psi}(y)]=0,  \label{eq:(2.8)}
\end{equation}
\begin{equation}
\{{\psi}_{\alpha}(x),{\psi}_{\beta}(y)\}=0, \quad 
\{{\psi}_{\alpha}(x),{\psi}_{\beta}^{\ast}(y)\}=(({\gamma}^0{\gamma}^-)^{-1})
_{{\alpha}{\beta}}{\delta}(x^+-y^+).  \label{eq:(2.9)}
\end{equation}

We begin constructing an operator solution by imposing the chiral anomaly
\begin{equation}
{\partial}_-J_+-{\partial}_+J_-=-m^2F_{-+}.  \label{eq:(2.10)}
\end{equation}
The current operator given by (\ref{eq:(2.2)}) is
\begin{equation}
J_-=g_{-+}J^++g_{--}J^-=-{\partial}^+F_{-+}-n_-B, \quad 
J_+={\partial}^-F_{-+}-n_+B.  \label{eq:(2.11)}
\end{equation}
Substituting these expressions into (\ref{eq:(2.10)}) yields
\begin{equation}
({\partial}_{\mu}{\partial}^{\mu}+m^2)F_{-+}=-n_-{\partial}_+B. 
\label{eq:(2.12)}
\end{equation}
Because ${\partial}_-B=0$, $F_{-+}$ is solved to be
\begin{equation}
F_{-+}={\partial}_-A_+=-m\tilde{\Sigma}-\frac{n_-}{m^2-n_-{\partial}_+^2}
{\partial}_+B. \label{eq:(2.13)}
\end{equation}
Here we have identified the homogeneous solution with the massive free field 
$\tilde{\Sigma}$, since $F_{-+}$ is gauge invariant. The normalization is 
determined in such a way that it agrees with the Landau gauge solution$^{18)}$. $\tilde{\Sigma}$ is given by
\begin{equation}
\tilde{\Sigma}(x)=\frac{1}{\sqrt{4{\pi}}}\int^{\infty}_{-{\infty}}\frac{dp_+}
{\sqrt{p^-}}\{ a(p_+)e^{-ip{\cdot}x}+a^{\ast}(p_+)e^{ip{\cdot}x} \},
\end{equation}
where $p^-=\sqrt{p_+^{\;\;2}+m_0^{\;\;2}}$ with $m_0^{\;\;2}=n_-m^2$ and 
\begin{equation}
[a(p_+),a(q_+)]=0, \quad [a(p_+),a^{\ast}(q_+)]={\delta}(p_+-q_+).
\end{equation}

By integrating (\ref{eq:(2.13)}) with respect to $x^-$ we obtain the general 
solution for $A_+$ as    
\begin{equation}
A_+=-\frac{m}{{\partial}_-}\tilde{\Sigma}-\frac{n_-x^-{\partial}_+}
{m^2-n_-{\partial}_+^{\;\;2}}B+{\rm integration\; constant}. \label{eq:(2.14)}
\end{equation}
We determine the integration constant in the following way: From the form of the d'Alembertian in the auxiliary coordinates,
\begin{equation}
{\partial}_{\mu}{\partial}^{\mu}=n_-{\partial}_-^{\;\;2}+2n_+n_-{\partial}_+
{\partial}_--n_-{\partial}_+^{\;\;2}  \label{eq:(2.15)}
\end{equation} 
we rewrite the first term of (\ref{eq:(2.14)}) as
\begin{equation}
\frac{1}{{\partial}_-}\tilde{\Sigma}=-\frac{n_+{\partial}_++{\partial}^-}
{m^2-n_-{\partial}_+^{\;\;2}}\tilde{\Sigma}.  \label{eq:(2.16)}
\end{equation} 
In the range of $\theta$ we are considering, we see that the operator $({\partial}_-)^{-1}$ does not give rise to any infrared divergences. Now the first term in (\ref{eq:(2.14)}) has the property $[\frac{m}{{\partial}_-}\tilde{\Sigma}(x^+),\frac{m}{{\partial}_-}\tilde{\Sigma}(y^+)]\neq 0$ at equal $x^-$-time because the $\tilde
{\Sigma}$ satisfies the following equal $x^-$-time commutation relations
\begin{equation}
[\tilde{\Sigma}(x),\tilde{\Sigma}(y)]=
[{\partial}^-\tilde{\Sigma}(x),{\partial}^-\tilde{\Sigma}(y)]=0, \; 
[\tilde{\Sigma}(x),{\partial}^-\tilde{\Sigma}(y)]=i{\delta}(x^+-y^+).
\label{eq:(2.17)}
\end{equation}
Therefore, to preserve the canonical commutation relations (\ref{eq:(2.7)}), we must
cancel the nonvanishing terms by contributions from another field $C$, the integration constant, which depends only on $x^+$. We find that if $B$ and $C$ satisfy the commutation 
relations
\begin{eqnarray}
&[B(x),B(y)]=[C(x),C(y)]=0, \; 
[B(x),C(y)]=i(m^2-n_-{\partial}_+^{\;\;2}){\delta}(x^+-y^+),& \nonumber \\
&[B(x),\tilde{\Sigma}(y)]=[C(x),\tilde{\Sigma}(y)]=0,& \label{eq:(2.18)}
\end{eqnarray}
then 
\begin{equation}
A_+=-\frac{m}{{\partial}_-}\tilde{\Sigma}+\frac{{\partial}_+}
{m^2-n_-{\partial}_+^{\;\;2}}\left(C-n_-x^-B\right)+\frac{n_+m^2}
{(m^2-n_-{\partial}_+^{\;\;2})^2}B  \label{eq:(2.19)}
\end{equation}
satisfies the quantization conditions in (\ref{eq:(2.7)}). In this way we 
can determine the integration constant: by taking account of the canonical 
quantization conditions in the temporal gauge formulation.

Now that we have the $A_+$, we proceed to solve the fermion field 
equation (\ref{eq:(2.3)}). A formal solution is given by
\begin{equation}
{\psi}_{\alpha}(x)=\frac{Z_{\alpha}}{\sqrt{({\gamma}^0{\gamma}^-)_
{{\alpha}{\alpha}}}}
{\rm exp}[-i\sqrt{\pi}{\Lambda}_{\alpha}(x)],  \label{eq:(2.20)}
\end{equation}
where $Z_{\alpha}$ is normalization constant and 
\begin{equation}
{\Lambda}_{\alpha}(x)=X(x)+(-1)^{\alpha}{\lambda}(x) \label{eq:(2.21)}
\end{equation}
with
\begin{eqnarray}
X&=&\frac{{\partial}^+}{{\partial}_-}\tilde{\Sigma}+
\frac{m}{m^2-n_-{\partial}_+^{\;\;2}}\left(C-n_-x^-B+\frac{n_+n_-}
{m^2-n_-{\partial}_+^{\;\;2}}{\partial}_+B\right), \nonumber \\
{\lambda}&=&\tilde{\Sigma}+\frac{m}{m^2-n_-{\partial}_+^{\;\;2}}
{\partial}_+^{-1}B. \label{eq:(2.22)}
\end{eqnarray}
Here the operator ${\partial}_+^{-1}$ is defined by
\begin{equation}
{\partial}_+^{-1}B(x)=\frac{1}{2}\int_{-{\infty}}^{\infty}dy^+
{\varepsilon}(x^+-y^+)B(y), \label{eq:(2.23)}
\end{equation} 
which imposes, in effect, the principal value regularization. To further examine the properties of the fermion field operators, such as their vacuum expectation 
values, we need explicit representations of the $B$ and $C$. Note that the canonical commutation relations cannot uniquely determine the representations of the canonical variables, as was pointed out by Haller$^{19)}$. One possibility is to choose the following zero norm and positive metric expressions for $B$ and $C$: 
\begin{eqnarray}
B(x)&=&\frac{m}{\sqrt{4\pi}}\int^{\infty}_{-\infty}dk_+\sqrt{|k_+|}\{ (a_0(k_+)
-a_1(k_+))e^{-ik_+x^+} + {\rm h.c.} \}, \nonumber \\ 
\frac{m}{m^2-n_-{\partial}_+^{\;\;2}}C(x)&=&\frac{i}{\sqrt{4\pi}}\int^{\infty}_
{-\infty}\frac{dk_+}{\sqrt{|k_+|}}\{ a_1(k_+)e^{-ik_+x^+} - a_1^{\ast}(k_+)
e^{ik_+x^+} \}, \label{eq:(2.24)}
\end{eqnarray}
where 
\begin{equation}
[a_0(k_+),a_0^{\ast}(q_+)]=-{\delta}(k_+-q_+), \quad 
[a_1(k_+),a_1^{\ast}(q_+)]={\delta}(k_+-q_+), \label{eq:(2.25)}
\end{equation}
and all other commutators are zero. It can be shown that $B$ and $C$ in 
(\ref{eq:(2.24)}) realize the PV prescription. With this choice, the spurion 
operators are composed out of ultraviolet parts of the $\tilde{\Sigma}$ as well as infrared 
parts of $C$ and ${\partial}_+^{-1}B$ according to the method given in 
Ref.18). This in turn gives rise to vanishing vacuum expectation values for 
the fermion field operators. In the Landau gauge solution$^{18)}$ the fermion 
fields have vanishing vacuum expectation value but in the light-cone gauge 
solution the vacuum expectation value of $\psi_+$ diverges.  The inclusion of 
the ultraviolet parts of $\tilde{\Sigma}$ in the spurions causes them to 
become $x$ dependent and the solution has some different characteristics from 
the previous solutions.  We are not completely certain whether the solution 
using the choice (\ref{eq:(2.24)}) is completely consistent but we do know 
that when we move to the axial gauge choice ($\theta > \frac{\pi}{4}$) we 
cannot use the representation (\ref{eq:(2.24)}): if we do there are 
uncancelled infrared divergences which destroy the solution. We will not 
further discuss the choice (\ref{eq:(2.24)}) in the present paper but will 
concentrate on another choice, one which realizes the ML prescription.  To do 
that we must limit the $k_+$-integration region to be $(0,\infty)$; we showed 
in Ref.13) that by choosing a suitable vacuum it is always possible to do so.  
We show in the next section that with the new choice, the infrared divergences 
in the axial gauge cancel.

To implement the new(ML) representation we choose the following representation for $B$ and 
$C$:
\begin{eqnarray}
B(x)&=&\frac{m}{\sqrt{2\pi}}\int^{\infty}_0dk_+\sqrt{k_+}\{ B(k_+)e^{-ik_+x^+} 
+ B^{\ast}(k_+)e^{ik_+x^+} \}, \nonumber \\ 
\frac{m}{m^2-n_-{\partial}_+^{\;\;2}}C(x)&=&\frac{i}{\sqrt{2\pi}}\int^{\infty}_
0\frac{dk_+}{\sqrt{k_+}}\{ C(k_+)e^{-ik_+x^+} - C^{\ast}(k_+)
e^{ik_+x^+} \}, \label{eq:(2.26)}
\end{eqnarray}
where 
\begin{equation}
[B(k_+),C^{\ast}(q_+)]=[C(k_+),B^{\ast}(q_+)]=-{\delta}(k_+-q_+),  
\label{eq:(2.27)}
\end{equation}
and all other commutators are zero. Note that the $C$ becomes a zero norm 
field as a consequence of altering the representations. Note also that the vacuum expectation value of ${\psi}_2$ diverges as it does in the light-cone gauge solution$^{9)}$. 

We define the physical subspace, $V$, by
\begin{equation}
V=\{\;\;|{\rm phys}{\rangle}\;\;|\;\; B(k_+)|{\rm phys}{\rangle}=0\;\; \}.
\end{equation}
To subtract the infrared parts of the massless fields into the spurion operators ${\sigma}_{\alpha}$, we define the infrared parts,
${\Lambda}_{\alpha}^{(0)}$, of ${\Lambda}_{\alpha}$ to be 
\begin{equation}
{\Lambda}_{\alpha}^{(0)}=\frac{i}{\sqrt{2{\pi}}}\int_0^{\kappa}\frac{dk_+}
{\sqrt{k_+}}\{C(k_+)-C^{\ast}(k_+)+(-1)^{\alpha}(B(k_+)-B^{\ast}(k_+))
\} \label{eq:(2.29)}
\end{equation}
where ${\kappa}$ is a small positive constant. Then the fermion field 
operators are defined as
\begin{equation}
{\psi}_{\alpha}(x)=\frac{Z_{\alpha}}{\sqrt{({\gamma}^0{\gamma}^-)_
{{\alpha}{\alpha}}}}
{\rm exp}[-i\sqrt{\pi}{\Lambda}_{{\alpha}r}^{(-)}(x)]{\sigma}_{\alpha}
{\rm exp}[-i\sqrt{\pi}{\Lambda}_{{\alpha}r}^{(+)}(x)]  \label{eq:(2.30)}
\end{equation}
where ${\Lambda}_{{\alpha}r}^{(-)}$ and ${\Lambda}_{{\alpha}r}^{(+)}$  
are the creation and annihilation operator parts of ${\Lambda}_{{\alpha}r}{\equiv}
{\Lambda}_{\alpha}-{\Lambda}_{\alpha}^{(0)}$ and
\begin{equation}
{\sigma}_{\alpha}={\rm exp}\left[-i\sqrt{\pi}\left({\Lambda}^{(0)}_{\alpha}
-(-1)^{\alpha}\frac{Q}{2m}\right)\right] \label{eq:(2.31)}
\end{equation}
Here, $Q=-\int^{\infty}_{-{\infty}}dx^+B(x)$; note that $Q$ in ${\sigma}_{\alpha}$ constitutes a Klein transformation.  

We enumerate properties of the ${\psi}_{\alpha}$ to show that the ${\psi}_
{\alpha}$ constitute the operator solution of the temporal gauge Schwinger 
model in the auxiliary coordinates.

(1) The Dirac equation is satisfied:
\begin{equation}
i{\gamma}^{\mu}({\partial}_{\mu}+ieA_{\mu}){\psi}=0  \label{eq:(2.32)}
\end{equation} 
To get this we use the fact that $[A_+^{(+)}(x),{\psi}_{\alpha}(x)]=0$, where $A_+^{(+)}$ is the annihilation operator part of $A_+$.

(2) The canonical commutation relations (\ref{eq:(2.8)}) and 
anticommutation relations (\ref{eq:(2.9)}) are satisfied.

(3) By applying the gauge invariant point-splitting procedure to e$\bar
{\psi}{\gamma}^{\mu}{\psi}$, we obtain the vector current $J^{\mu}=m
{\varepsilon}^{{\mu}{\nu}}{\partial}_{\nu}{\lambda}$, where
${\varepsilon}^{++}={\varepsilon}^{--}=0,\;{\varepsilon}^{+-}=
-{\varepsilon}^{-+}=1$. 
The charge operator, $Q$, is given by
\begin{equation}
Q=\int^{\infty}_{-{\infty}}dx^+J^-(x)=-\int^{\infty}_{-{\infty}}dx^+B(x), 
  \label{eq:(2.33)}
\end{equation}
where the derivative terms integrate to zero.

(4) The conserved chiral current $J^{{\mu}5}=m{\varepsilon}^{{\mu}{\nu}}
{\partial}_{\nu}X$ is similarly obtained from e$\bar{\psi}{\gamma}^{\mu}
{\gamma}^5{\psi}$. The chiral charge operator, $Q_5$, is given by
\begin{equation}
Q_5=-\int^{\infty}_{-{\infty}}dx^+J^{-5}(x)
=\int^{\infty}_{-{\infty}}dx^+{\partial}_+C(x), 
  \label{eq:(2.34)}
\end{equation}
where the derivative terms, except ${\partial}_+C $, integrate to zero. The 
term  ${\partial}_+C $ does not integrate to zero because ${\psi}$ contains 
the singular operator ${\partial}_+^{-1}B$.

(5) Applying the gauge invariant point-splitting procedure to the fermi products in the symmetric energy-momentum tensor and subtracting a diverging c-number, which
we denote by $R$, we get
\begin{eqnarray}
{\Theta}_+^{\;\;-}&=&\frac{i}{2}R(\bar{\psi}{\gamma}^-{\partial}_+{\psi}
-{\partial}_+\bar{\psi}{\gamma}^-{\psi})-A_+J^--\frac{1}{2}(A_+,B)_+ 
\nonumber \\
&=&\frac{1}{2}({\partial}_+{\lambda},{\partial}^-{\lambda})_+
-\frac{1}{2}(A_+,B)_+,  \label{eq:(2.35)}
\end{eqnarray}
\begin{eqnarray}
{\Theta}_+^{\;\;+}&=&\frac{i}{2}R(\bar{\psi}{\gamma}^+{\partial}_+{\psi}
-{\partial}_+\bar{\psi}{\gamma}^+{\psi})-A_+J^+ +\frac{1}{2}(F_{-+})^2 
\nonumber \\
&=&-\frac{n_-}{2}\{({\partial}_+{\lambda})^2+({\partial}_-{\lambda})^2 \}
+\frac{1}{2}(F_{-+})^2,  \label{eq:(2.36)}
\end{eqnarray}
\begin{eqnarray}
{\Theta}_-^{\;\;-}&=&-\frac{i}{2}R(\bar{\psi}{\gamma}^+{\partial}_+{\psi}
-{\partial}_+\bar{\psi}{\gamma}^+{\psi})+A_+J^+ +\frac{1}{2}(F_{-+})^2 
\nonumber \\
&=&\frac{n_-}{2}\{({\partial}_+{\lambda})^2+({\partial}_-{\lambda})^2 \}
+\frac{1}{2}(F_{-+})^2,  \label{eq:(2.37)}
\end{eqnarray}
\begin{eqnarray}
{\Theta}_-^{\;\;+}&=&-\frac{in_+}{2n_-}R(\bar{\psi}{\gamma}^+{\partial}_+
{\psi}-{\partial}_+\bar{\psi}{\gamma}^+{\psi}) 
-\frac{i}{2n_-}R(\bar{\psi}{\gamma}^+{\gamma}^5{\partial}_+{\psi}
-{\partial}_+\bar{\psi}{\gamma}^+{\gamma}^5{\psi}) \nonumber \\
&+&\frac{1}{n_-}J_-A_+ +\frac{n_+}{n_-}J^+A_+ \quad 
=\frac{1}{2}({\partial}_-{\lambda},{\partial}^+{\lambda})_+.
 \label{eq:(2.38)}
\end{eqnarray}

(6) Translational generators consist of those of the constituent fields: 
\begin{eqnarray}
P_-&=&\int^{\infty}_{-{\infty}}dx^+:{\Theta}_-^{\;\;-}: \nonumber \\
&=&\int^{\infty}_{-{\infty}}dx^+\frac{1}{2}:
\{ n_-({\partial}_+\tilde{\Sigma})^2+n_-({\partial}_-\tilde{\Sigma})^2
+m^2(\tilde{\Sigma})^2+B\frac{n_-}
{m^2-n_-{\partial}_+^{\;\;2}}B \}:, \nonumber \\
P_+&=&\int^{\infty}_{-{\infty}}dx^+:{\Theta}_+^{\;\;-}:
=\int^{\infty}_{-{\infty}}dx^+:\{ {\partial}_+\tilde{\Sigma}
{\partial}^-\tilde{\Sigma}-B\frac{1}{m^2-n_-{\partial}_+^{\;\;2}}{\partial}_+C 
\}:. \label{eq:(2.39)}
\end{eqnarray}

\section{Cancelation of Infrared divergences resulting from ${\partial}_-^{-1}$}

In this section we continue ${\theta}$ into the region $\frac{\pi}{4}<
{\theta}<\frac{\pi}{2}$ and  take $x^+$ as the evolution parameter. In 
accordance with the change of the evolution parameter we change the Fourier 
expansion of $\tilde{\Sigma}$ so as to be convenient for calculating equal 
$x^+$-time commutation relations. By making use of $\frac{dp_+}{p^-}=-\frac
{dp_-}{p^+}$ we change integration variables from $p_+$ to $p_-$ and 
rewrite $\tilde{\Sigma}$ as follows
\begin{equation}
\tilde{\Sigma}(x)=\frac{1}{\sqrt{4{\pi}}}\int^{\infty}_{-{\infty}}\frac{dp_-}
{\sqrt{p^+}}\{ a(p_-)e^{-ip{\cdot}x}+a^{\ast}(p_-)e^{ip{\cdot}x} \},
\end{equation}
where $p^+=\sqrt{p_-^{\;\;2}+m_0^{\;\;2}}$ with $m_0^{\;\;2}=|n_-|m^2$ and 
\begin{equation}
[a(p_-),a(q_-)]=0, \quad [a(p_-),a^{\ast}(q_-)]={\delta}(p_--q_-).
\end{equation}
$\tilde{\Sigma}$ satisfies the following equal 
$x^+$-time commutation relations:
\begin{equation}
[\tilde{\Sigma}(x),\tilde{\Sigma}(y)]=
[{\partial}^+\tilde{\Sigma}(x),{\partial}^+\tilde{\Sigma}(y)]=0, \; 
[\tilde{\Sigma}(x),{\partial}^+\tilde{\Sigma}(y)]=i{\delta}(x^--y^-).
\label{eq:(3.3)}
\end{equation}
Furthermore, in accordance with the change of the evolution parameter, the
conjugate momentum of ${\psi}$ changes to
\begin{equation}
{\pi}_{\psi}^{(a)}=\frac{{\delta}L}{{\delta}{\partial}_+{\psi}}=
i\bar{\psi}{\gamma}^+. \label{eq:(3.4)}
\end{equation}
Therefore, we have to modify the normalization of ${\psi}$ in (\ref{eq:(2.30)})
accordingly:
\begin{equation}
{\psi}_{\alpha}^{(a)}(x)=\frac{Z_{\alpha}^{(a)}}{\sqrt{({\gamma}^0{\gamma}^+)_
{{\alpha}{\alpha}}}}
{\rm exp}[-i\sqrt{\pi}{\Lambda}_{{\alpha}r}^{(-)}(x)]{\sigma}_{\alpha}
{\rm exp}[-i\sqrt{\pi}{\Lambda}_{{\alpha}r}^{(+)}(x)],  \label{eq:(3.5)}
\end{equation}
where we have put $(a)$ on ${\psi}$ and on the normalization constant to 
denote that the normalization is altered.

We begin by pointing out that because $n_-=\cos 2{\theta}<0$ in the region  
$\frac{\pi}{4}<{\theta}<\frac{\pi}{2}$, the Laplace operator, $m^2-n_-{\partial}_+^
{\;\;2}$, which operates on the residual gauge fields, becomes hyperbolic so 
that its inverse gives rise to divergences. Therefore, we regularize them by 
PV regularization. We then recover well-defined fermion field operators because linear infrared divergences resulting from $\frac{{\partial}_+}{{\partial}_-}\tilde{\Sigma}$ are canceled by those resulting from the square of the inverse of the hyperbolic Laplace operator. 
To show that this really happens, we rewrite ${\psi}^{(a){\ast}}_{\alpha}(x)
{\psi}_{\alpha}^{(a)}(y)$ into  normal ordered form. It is tedious but 
straightforward to obtain
\begin{eqnarray}
{\psi}^{(a){\ast}}_{\alpha}(x){\psi}_{\alpha}^{(a)}(y)&=&
\frac{(Z_{\alpha}^{(a)})^2}{({\gamma}^0{\gamma}^+)_{{\alpha}{\alpha}}}
{\rm exp}[M_{{\alpha}{\alpha}}(x-y)] 
{\rm exp}[-i\sqrt{\pi}({\Lambda}_{\alpha}
^{(-)}(x)-{\Lambda}_{\alpha}^{(-)}(y))]  \nonumber \\
&{\times}&{\rm exp}[-i\sqrt{\pi}({\Lambda}_
{\alpha}^{(+)}(x)-{\Lambda}_{\alpha}^{(+)}(y))], \label{eq:(3.6)}
\end{eqnarray}
where
\begin{eqnarray}
&M_{{\alpha}{\alpha}}(x)=\frac{1}{4}\int^{\infty}_{-{\infty}}\frac{dp_-}{p^+}
\left( 1+(\frac{p^+}{p_-})^2+2(-1)^{\alpha}\frac{p^+}{p_-} \right)
e^{-ip{\cdot}x}& \nonumber \\ 
&+\frac{1}{2}\int^{\infty}_0dk_+\left(-\frac{im^2n_-x^-}{m^2+n_-k_+^{\;\;2}}
+\frac{2m^2n_-n_+k_+}{(m^2+n_-k_+^{\;\;2})^2} \right)e^{-ik_+x^+}& \nonumber \\
&+(-1)^{\alpha}\int^{\infty}_0dk_+\left(\frac{n_-k_+}{m^2+n_-k_+^{\;\;2}}
e^{-ik_+x^+}-\frac{e^{-ik_+x^+}-{\theta}({\kappa}-k_+)}{k_+} \right).& 
\label{eq:(3.7)}
\end{eqnarray}
We must show that $M_{{\alpha}{\alpha}}(x)$ does not diverge when 
$x^+=0$, because in what follows  we make use of the values of $M_{{\alpha}
{\alpha}}(x-y)$ at $x^+=y^+$ to calculate the equal $x^+$-time anticommutation 
relations and to calculate the dynamical operators. 

Since $(p^+)^2=p_-^{\;\;2}+m_0^{\;\;2}$, only the term $\frac{m_0^{\;\;2}}
{p_-^{\;\;2}}$  among the physical contributions on the first line of 
(\ref{eq:(3.7)}) gives rise to a linear infrared divergence. We rewrite the expression as follows
\begin{equation}
\frac{1}{4}\int^{\infty}_{-{\infty}}\frac{dp_-}{p^+}\frac{m_0^{\;\;2}}
{p_-^{\;\;2}}e^{-ip_-x^-}
=\int^{\infty}_0\frac{dq}{q^2}-\frac{m_0^{\;\;2}}{2}
\int^{\infty}_0\frac{dp_-}{p^+}\frac{1-\cos p_-x^-}{p_-^{\;\;2}}, 
\label{eq:(3.8)}
\end{equation}
where the first, linearly diverging integration is obtained 
by changing integration variables from $p_-$ to $q$ through $p_-=\frac{m_0q}
{2\sqrt{q+1}}$.

As for the residual gauge field's contributions: those divergences which can be regularized by the PV prescription we regularize by the PV prescription. Then we find that when $x^+=0$, the first term on the 
second line of (\ref{eq:(3.7)}) is equal to zero:
\begin{equation}
\int_0^{\infty}dk_+\frac{n_-m^2}{m^2+n_-k_+^{\;\;2}}=
\int_0^{\infty}dk_+\frac{m^2}{k_+^{\;\;2}-a^2}=0, \label{eq:(3.9)}
\end{equation}
where $a=\frac{m}{\sqrt{-n_-}}$. However, we cannot regularize a linear 
divergence resulting from a double pole by the PV regularization. Therefore 
when $x^+=0$, the second term on the second line of (\ref{eq:(3.7)}) gives 
rise to a linear divergence as follows 
\begin{eqnarray}
&{}&\int_0^{\infty}dk_+\frac{n_-n_+m^2k_+}{(m^2+n_-k_+^{\;\;2})^2}=
-\frac{n_+a}{4}\int_0^{\infty}dk_+\left(\frac{1}{(k_+-a)^2}-
\frac{1}{(k_++a)^2}\right) \nonumber \\
&=& -\frac{n_+a}{4}\left(\int_0^adk\frac{1}{k^2}+\int_0^{\infty}dk
\frac{1}{k^2}-\int_a^{\infty}dk\frac{1}{k^2} \right) =-\int_0^{\infty}dq
\frac{1}{q^2}+\frac{n_+}{2}, \label{eq:(3.10)}
\end{eqnarray} 
where the last, linearly diverging term is given by changing integration 
variables from $k$ to $q$ through $k=\frac{n_+a}{2}q$. Thus we see that the 
linear divergence originated from the physical contributions is canceled 
by that resulting from the residual gauge field's contributions. It should be 
noted here that if we had chosen the representation of the residual gauge 
fields so as to realize the PV prescription (the expression given by (\ref{eq:(2.24)})), we would obtain the identically vanishing integral $
\int_{-{\infty}}^{\infty}dk_+\frac{n_-n_+m^2k_+}{(m^2+n_-k_+^{\;\;2})^2}=0
$ so that the linear divergence resulting from 
$\frac{{\partial}^+}{{\partial}_-}\tilde{\Sigma}$ is not canceled.

It turns out that when $x^+=0$, the terms on the third line of 
(\ref{eq:(3.7)}) give rise to a finite value, $\log \frac{\kappa}
{\sqrt{|{\kappa}^2-a^2|}}$, so that by summing all the  finite contributions we obtain
\begin{eqnarray}
&M_{{\alpha}{\alpha}}(x)|_{x^+=0}=K_0(m_0|x^-|)+i(-1)^{{\alpha}-1}
\frac{\pi}{2}{\varepsilon}(x^-)& \nonumber \\
&-\frac{m_0^{\;\;2}}{2}
\int^{\infty}_0\frac{dp_-}{p^+}\frac{1-\cos p_-x^-}{p_-^{\;\;2}} +\frac{n_+}{2}
+(-1)^{\alpha}\log \frac{\kappa}{\sqrt{|{\kappa}^2-a^2|}}& \nonumber \\
&=-\left(\log \frac{m_0{\gamma}}{2} +\log (x^-+i{\varepsilon}(-1)^{{\alpha}-1})
\right)+i(-1)^{{\alpha}-1}\frac{\pi}{2} +F_{\alpha}(|x^-|),&
  \label{eq:(3.11)}
\end{eqnarray}
where $K_0(m_0|x^-|)$ is the modified Bessel function of order 0, which possesses  
a logarithmic term: $-( \log \frac{m_0|x^-|}{2} +{\gamma})$. We have gathered the
remainder of $K_0$ and the other terms to make a continuous function 
$F_{\alpha}(|x^-|)$.

Now that we have finished showing that the linear divergences cancel each 
other, we proceed to study other properties of the fermion field operators.

(1) The equal $x^+$-time anticommutation relations hold. For example, when $x^+=
y^+$, we find that 
\begin{equation}
{\rm exp}[M_{{\alpha}{\alpha}}(x-y)]=\frac{2}{m_0e^{\gamma}}
\frac{i(-1)^{{\alpha}-1}}{x^--y^-+i{\varepsilon}(-1)^{{\alpha}-1}}
e^{F_{\alpha}(|x^--y^-|)},
\end{equation} 
so that if we take $Z_{\alpha}^{(a)}$ to be
\begin {equation}
(Z_{\alpha}^{(a)})^2=\frac{m_0e^{\gamma}}{4{\pi}}e^{-F_{\alpha}(0)},
\end{equation}
then  we obtain
\begin{eqnarray}
&{}&\{{\psi}^{\ast}_{\alpha}(x),{\psi}_{\alpha}(y)\}|_{x^+=y^+} \nonumber \\
&=&\frac{{\rm exp}[F_{\alpha}(|x^--y^-|)-F_{\alpha}(0)]}{2{\pi}({\gamma}^0
{\gamma}^+)_{{\alpha}{\alpha}}}
\left(\frac{i(-1)^{{\alpha}-1}}{x^--y^-+i{\varepsilon}(-1)^{{\alpha}-1}}
+\frac{i(-1)^{{\alpha}-1}}{y^--x^-+i{\varepsilon}(-1)^{{\alpha}-1}}\right) 
\nonumber \\
&{\times}&{\rm exp}[-i\sqrt{\pi}({\Lambda}_{\alpha}
^{(-)}(x)-{\Lambda}_{\alpha}^{(-)}(y))]  
{\rm exp}[-i\sqrt{\pi}({\Lambda}_
{\alpha}^{(+)}(x)-{\Lambda}_{\alpha}^{(+)}(y))] \nonumber \\
&=&\frac{1}{({\gamma}^0{\gamma}^+)_{{\alpha}{\alpha}}}{\delta}(x^--y^-). 
\label{eq:(3.14)}
\end{eqnarray}
When ${\alpha}{\ne}{\beta}, \{{\psi}^{\ast}_{\alpha}(x),{\psi}_{\beta}(y)\}
|_{x^+=y^+}$ vanishes due to the Klein transformation factors.  

(2) The vector current, $J^{\mu}=m{\varepsilon}^{{\mu}{\nu}}{\partial}_{\nu}
{\lambda}$, is obtained this time by applying the equal $x^+$-time point-splitting 
procedure to e$\bar{\psi}{\gamma}^{\mu}{\psi}$.

(3) The conserved chiral current $J^{{\mu}5}=m{\varepsilon}^{{\mu}{\nu}}
{\partial}_{\nu}X$ is similarly obtained from e$\bar{\psi}{\gamma}^{\mu}
{\gamma}^5{\psi}.$

(4) As a consequence of our choice of representation of the residual 
gauge fields, we obtain the ML 
form of the $x^+$-ordered gauge field propagator:
\begin{eqnarray}
D_{{\mu}{\nu}}(x-y)&=&{\langle}0|\{{\theta}(x^+-y^+)A_{\mu}(x)
A_{\nu}(y)+{\theta}(y^+-x^+)A_{\nu}(y)A_{\mu}(x) \}|0{\rangle}
 \nonumber \\
&=&\frac{1}{(2{\pi})^2}\int d^2qD_{{\mu}{\nu}}(q)e^{-iq \cdot (x-y)}
\label{eq:(3.15)} 
\end{eqnarray}
where
\begin{eqnarray}
D_{{\mu}{\nu}}(q)&=&\frac{i}{q^2-m^2+i{\epsilon}} \left( -g_{{\mu}{\nu}}
+\frac{n_{\mu}q_{\nu}+n_{\nu}q_{\mu}}{q_-+i{\varepsilon}{\rm sgn}(q_+)}-
n^2\frac{q_{\mu}q_{\nu}}{(q_-+i{\varepsilon}{\rm sgn}(q_+))^2 } 
\right) \nonumber \\
&-&{\delta}_{{\mu}+}{\delta}_{{\nu}+}\frac{i}{2}
\left(\frac{1}{(q_-+i{\varepsilon})^2}+\frac{1}{(q_-+i{\varepsilon})^2}\right).
\label{eq:(3.16)} 
\end{eqnarray}
We give a detailed derivation of this in the Appendix. 

(5) Applying the gauge invariant point-splitting procedure to the fermi products in the symmetric energy-momentum tensor and subtracting a diverging c-number, 
which we denote by $R$, we get
\begin{eqnarray}
{\Theta}_+^{\;\;-}&=&\frac{in_+}{2n_-}R(\bar{\psi}{\gamma}^-{\partial}_-{\psi}
-{\partial}_-\bar{\psi}{\gamma}^-{\psi})
-\frac{i}{2n_-}R(\bar{\psi}{\gamma}^-{\gamma}^5{\partial}_-{\psi}
-{\partial}_-\bar{\psi}{\gamma}^-{\gamma}^5{\psi}) \nonumber \\
&-&\frac{1}{2}(A_+,B)_+ \quad =\frac{1}{2}({\partial}_+{\lambda},{\partial}^-
{\lambda})_+-\frac{1}{2}(A_+,B)_+,  \label{eq:(3.17)}
\end{eqnarray}
\begin{eqnarray}
{\Theta}_+^{\;\;+}&=-&\frac{i}{2}R(\bar{\psi}{\gamma}^-{\partial}_-{\psi}
-{\partial}_-\bar{\psi}{\gamma}^-{\psi}) +\frac{1}{2}(F_{-+})^2 
\nonumber \\
&=&-\frac{n_-}{2}\{({\partial}_+{\lambda})^2+({\partial}_-{\lambda})^2 \}
+\frac{1}{2}(F_{-+})^2,  \label{eq:(3.18)}
\end{eqnarray}
\begin{eqnarray}
&{}&{\Theta}_-^{\;\;-}=\frac{1}{2}(F_{-+})^2 +\frac{i}{4}R(\bar{\psi}{\gamma}^-
{\partial}_-{\psi}-{\partial}_-\bar{\psi}{\gamma}^-{\psi}) \nonumber \\
&-&\frac{n_+}{4}R(\bar{\psi}{\gamma}^-{\gamma}^5{\partial}_-{\psi}
-{\partial}_-\bar{\psi}{\gamma}^-{\gamma}^5{\psi})
+\frac{n_-}{4}R(\bar{\psi}{\gamma}^+{\gamma}^5{\partial}_-{\psi}
-{\partial}_-\bar{\psi}{\gamma}^+{\gamma}^5{\psi}) \nonumber \\
&=&\frac{n_-}{2}\{({\partial}_+{\lambda})^2+({\partial}_-{\lambda})^2 \}
+\frac{1}{2}(F_{-+})^2,  \label{eq:(3.19)}
\end{eqnarray}
\begin{equation}
{\Theta}_-^{\;\;+}=\frac{i}{2}R(\bar{\psi}{\gamma}^+{\partial}_-{\psi}
-{\partial}_-\bar{\psi}{\gamma}^+{\psi}) 
=\frac{1}{2}({\partial}_-{\lambda},{\partial}^+{\lambda})_+.
 \label{eq:(3.20)}
\end{equation}

We must take special care in the axial gauge formulation when we derive 
the conserved translational generators from the divergence equation 
\begin{equation}
{\partial }_{\nu }{\Theta}_{{\mu}}^{\;\;{\nu }}=0.  \label{eq:(3.21)}
\end{equation}
The problem is that although $x^-$ is a space coordinate, ${\Theta}_{\mu}^
{\;\;-}$ does not vanish in the limits $x^-{\to}{\pm}{\infty}$. This is 
because the residual gauge fields do not depend on $x^-$ and because $A_+$ 
depends explicitly on $x^-$. Therefore we have to retain ${\lim}_{x^-\to
{\pm}{\infty}} {\Theta}_{\mu}^{\;\;-}$. To take this fact into account, we 
integrate the divergence equation ${\partial }_{\nu }{\Theta}_{{\mu}}^
{\;\;{\nu }}=0$ over a closed surface shown in Fig.~1, whose bounds $T$ and 
$L$ are let tend to ${\infty}$ after all calculations are finished. We remark 
that we can use this surface even in the limit ${\theta}\to\frac{\pi}{2}-0$, 
in contrast with one used in Refs. 8) and 20). 

\begin{figure}
\centerline{\psfig{figure=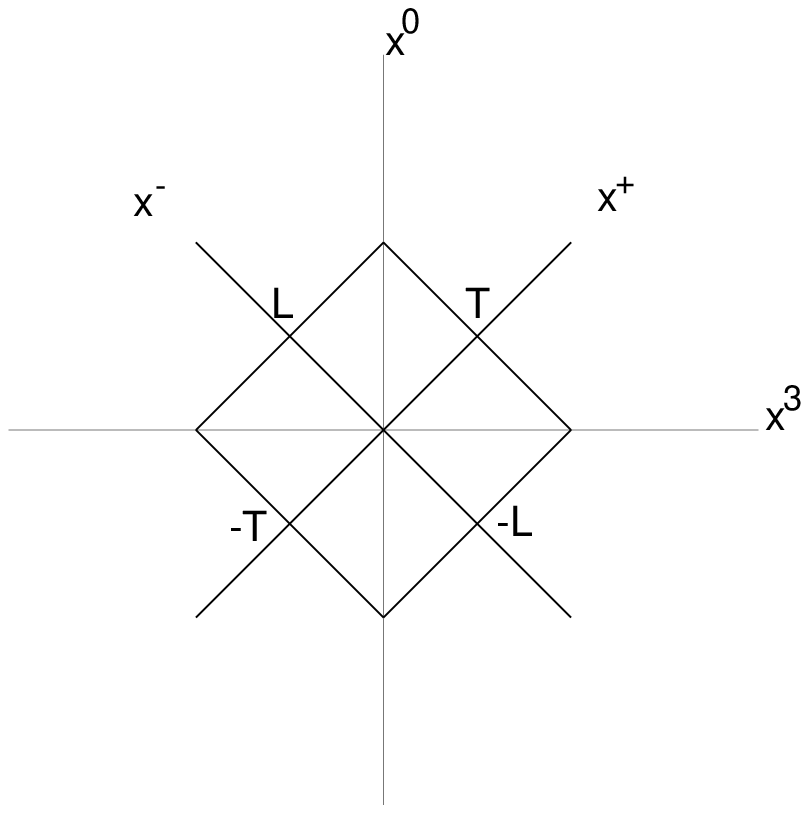,width=4.3in,height=4.3in}
}
\caption{}
\end{figure}

It is straightforward to obtain 
\begin{equation}
0= \left( {\int }_{-L}^{L}dx^- 
\left[{\Theta}_{\mu}^{\;\;+}(x) \right]^{x^+=T}_{x^+=-T}+
{\int }_{-T}^{T}dx^+ \left[{\Theta}_{\mu}^{\;\;-}
(x) \right]^{x^-=L}_{x^-=-L} \right),  \label{eq:(3.22)}
\end{equation}
where 
\begin{eqnarray}
\left[{\Theta}_{\mu}^{\;\;+}(x) \right]^{x^+=T}_{x^+=-T}&=&
{\Theta}_{\mu}^{\;\;+}(x)|_{x^+=T}-{\Theta}_{\mu}^{\;\;+}(x)|_{x^+=-T}, 
\nonumber \\
\left[{\Theta}_{\mu}^{\;\;-}(x) \right]^{x^-=L}_{x^-=-L}&=&
{\Theta}_{\mu}^{\;\;-}(x)|_{x^-=L}-{\Theta}_{\mu}^{\;\;-}(x)|_{x^-=-L}. 
\label{eq:(3.23)}
\end{eqnarray}
In what follows we refer to operators made of $\tilde{\Sigma}$ as physical 
operators and operators made of $B$ and $C$ as residual gauge operators.

It can be shown that products of physical operators and residual gauge 
operators vanish, so that the physical operator parts and the residual gauge 
operator parts decouple in (\ref{eq:(3.22)}). This result reflects the fact 
that the physical operator parts are conserved by themselves. It can be also 
shown that the residual gauge operators which are not well-defined in the 
limit $L{\to}{\infty}$ cancel among themselves. After getting 
the decoupled and well-defined expressions, we take the limit 
$L{\to}{\infty}$. This limit enables us to discard the physical operator parts 
${\theta}_{\mu}^{\;\;-}(x)$ of ${\Theta}_{\mu}^{\;\;-}(x)$, as is usually done 
when one calculates the conserved physical generators in the traditional axial 
gauge formulations. Finally, we take the limit $T{\to}{\infty}$. At this stage 
we assume that we can integrate the remaining residual gauge operators in 
${\Theta}_{\mu}^{\;\;-}(x)$ by parts in the $x^+$ direction. This assumption 
is justified because the residual gauge operator parts are conserved by 
themselves. As a consequence, we obtain
\begin{equation}
0=\int^{\infty}_{-{\infty}}dx^- \left[{\theta}_+^{\;\;+} \right]^
{x^+={\infty}}_{x^+=-{\infty}}
-\int^{\infty}_{-{\infty}}dx^+ \left[ B\frac{1}{m^2-n_-{\partial}_+^{\;\;2}}
{\partial}_+C \right]^{x^-={\infty}}_{x^-=-{\infty}},   
\label{eq:(3.24)}
\end{equation}
\begin{equation}
0=\int^{\infty}_{-{\infty}}dx^- \left[{\theta}_-^{\;\;+} \right]^
{x^+={\infty}}_{x^+=-{\infty}} 
+\frac{1}{2}{\int}^{\infty}_{-{\infty}}dx^+ \left[ B(x)\frac{n_-}{
m^2-n_-{\partial}_+^{\;\;2}}B(x)\right]^{x^-={\infty}}_{x^-=-{\infty}},  
\label{eq:(3.25)}
\end{equation}
where 
\begin{eqnarray}
{\theta}_+^{\;\;+}&=&\frac{n_+}{n_-}:{\partial}_-\tilde{\Sigma}
{\partial}^+\tilde{\Sigma}:
-\frac{1}{2n_-}:\{ ({\partial}^+\tilde{\Sigma})^2
 +({\partial}_-\tilde{\Sigma})^2 +m_0^{\;\;2}(\tilde{\Sigma})^2 \}:, 
\nonumber \\
{\theta}_-^{\;\;+}&=&:{\partial}_-\tilde{\Sigma}{\partial}^+\tilde{\Sigma}:.
\end{eqnarray}
>From these relations we obtain the conserved generators:
\begin{equation}
P_+=\int^{\infty}_{-{\infty}}dx^- {\theta}_+^{\;\;+}(x) 
-\int^{\infty}_{-{\infty}}dx^+  B(x)\frac{1}{m^2-n_-{\partial}_+^{\;\;2}}
{\partial}_+C(x),   
\label{eq:(3.26)}
\end{equation}
\begin{equation}
P_-=\int^{\infty}_{-{\infty}}dx^- {\theta}_-^{\;\;+}(x)  
+\frac{1}{2}\int^{\infty}_{-{\infty}}dx^+ B(x)\frac{n_-}
{m^2-n_-{\partial}_+^{\;\;2}}B(x).  \label{eq:(3.27)}
\end{equation}
In this way, even in the axial gauge formulation, we can obtain the conserved 
translational generators. These consist of the physical operator part 
integrated over $x^+=$ constant and the residual gauge operator part 
integrated over $x^-=$ constant. By making use 
of the commutation relations (\ref{eq:(2.18)}) and (\ref{eq:(3.3)}), it is 
easy to show that the Heisenberg equations for $A_+$ and ${\psi}^{(a)}$ hold. 
This shows that we have successfully constructed the extended Hamiltonian 
formulation of the axial gauge Schwinger model.

Finally, we shall discuss the charges in the axial gauge formulation.  The 
current, being gauge invariant, is the same as in the temporal gauge
\begin{equation}
J^-=-{\partial}_+(m\tilde{\Sigma}+\frac{m^2}{m^2-n_-{\partial}_+^2}
{\partial}_+^{-1}B), \;\; 
J^+={\partial}_-(m\tilde{\Sigma}+\frac{m^2}{m^2-n_-{\partial}_+^2}
{\partial}_+^{-1}B)=m{\partial}_-\tilde{\Sigma}. 
\end{equation}
Using the fact that the current has zero divergence and also using the relation
\begin{equation}
 \frac{m^2}{m^2-n_-{\partial}_+^2}B=B+
\frac{n_-{\partial}_+^2}{m^2-n_-{\partial}_+^2}B, 
\end{equation}
we find the charge to be
\begin{equation}
Q= -\int^{\infty}_{-{\infty}}dx^+B(x) + \int^{\infty}_{-{\infty}}dx^
-m{\partial}_-\tilde{\Sigma}(x).  \label{charge2}
\end{equation} 
In the temporal gauge formulation (and also in the light-cone gauge$^{9)}$  
and Landau gauge$^{18)}$  solutions) we set the second term equal to zero.  
The justification for that is that $\tilde{\Sigma}(x)$ is a massive field and 
since it has no infrared singularities, the term involving the integral over 
it commutes with the $\psi$ field.  The question is somewhat more complicated 
in the axial gauge case since the singularity coming from $\partial_-^{-1}$ 
causes the commutator of the second term in (\ref{charge2}) with fermi field 
to become nonzero.  On the other hand, we cannot keep both terms in 
(\ref{charge2}) since in that case the fermi field would carry twice the 
correct charge.  We believe that the correct charge operator is obtained from 
(\ref{charge2}) by letting the second term provide vanishing topological 
charges. In previous attempts to formulate the Coulomb gauge Schwinger model 
(using a representation space which is entirely physical), the fermion field 
operators were represented as solitons made solely out of the $\tilde{\Sigma}$
 so that  the $\tilde{\Sigma}$ tended to  nonvanishing values in the imits 
$x^- {\to}{\pm}{\infty}$. That led, inevitably, to infrared difficulties. In 
our formulation, the fermion field operators are constituted in the same 
manner as ones in the temporal gauge formulation so that it is reasonable for 
them not to carry any topological charges. To attain this we define the 
operator $({\partial}_-)^{-1}$ as a finite integral: 
\begin{equation}
\frac{1}{2}\int^L_{-L}{\varepsilon}(x^--y^-)f(y^-)dy^-
\end{equation}
and take the limit $L{\to}{\infty}$ after all relevant calculations are 
finished.  This procedure is justified by the fact that the infrared 
divergences are all eliminated. We therefore take the charge operator to be
\begin{equation}
Q= -\int^{\infty}_{-{\infty}}dx^+B(x). 
\end{equation} 
Similarly, we take the chiral charge to be
\begin{equation}
Q_5=\int^{\infty}_{-{\infty}}dx^+{\partial}_+C(x). 
\end{equation}

\section{Concluding remarks}

In this paper we have shown that $x^-$-independent residual gauge fields can 
be introduced as regulator fields in the pure space-like axial gauge 
formulation of the Schwinger model by extending the Hamiltonian formalism 
{\it \'a la} McCartor and Robertson. Because we do not have consistent pure 
space-like axial gauge quantization conditions, we have extrapolated 
the solution in the temporal gauge formulation and checked that it is also 
a solution in the axial gauge formulation. One conclusion to be drawn from the 
work involves the representation of the residual gauge fields:  We have found 
that in the axial gauge we must choose the representation which specifies the 
ML prescription in order for the linear infrared divergences resulting from 
the operator $({\partial}_-)^{-2}$ to be 
canceled. For this cancellation to work it is 
necessary for us to change integration variables to ones proportional to $n_+=
\sin 2{\theta}$. It follows from this that a pure space-like axial gauge 
formulation in the ordinary coordinates (this is the case of ${\theta}=\frac
{\pi}{2}$ which gives us the Coulomb gauge) has to be defined as the limit 
${\theta}{\to}\frac{\pi}{2}-0$. Regulating the solution in this way involves 
information off the initial value surface ($x^0 = 0$), but in any event the 
solution includes auxiliary fields and the representation space is of 
indefinite metric.  

In contrast, the case of ${\theta}=\frac{\pi}{4}$ (the light-cone axial 
gauge formulation) can be defined by simply setting ${\theta}=\frac{\pi}{4}$ 
(but the theory still has to be regulated by splitting the fermi products off 
the initial value surface).$^{9,16)}$ This is because, by virtue of $n^2=n_-=
\cos 2{\theta}=0$, we do not have the 
linear divergences resulting from $({\partial}_-)^{-2}$ and from the square of 
the inverse of the hyperbolic Laplace operator except for the contact term in 
the most singular component of the gauge field propagator. We have also found 
that in order for the equal $x^+$-time anticommutation relations to be 
satisfied in the axial gauges, we have to alter the representation of the 
$\tilde{\Sigma}$ from the temporal one to the axial one and make a 
modification to the overall normalization of the fermion field operators. 

In contrast to the axial gauges, it seems that in the temporal gauges we can 
use either the representation of the residual gauge fields which specifies the 
ML prescription or the one which specifies the PV prescription.  If we use the 
latter, then both components of the fermi field have vanishing vacuum 
expectation value. 

Now that we have the operator solution in the axial gauge Schwinger model, we 
can calculate commutation relations of the $A_{\mu}$. It turns out that
\begin{equation}
[A_{\mu}(x),A_{\nu}(y)]=i\{ -g_{{\mu}{\nu}}{\Delta}(x-y;m^2)+(n_{\mu}
{\partial}_{\nu}+n_{\nu}{\partial}_{\mu}){\partial}_-E(x-y)-n^2{\partial}_{\mu}
{\partial}_{\nu}E(x-y) \}, \label{eq:(4.1)}
\end{equation}
where ${\Delta}(x;m^2)$ is the commutator function of the free field of mass 
$m$ and 
\begin{equation}
E(x)=\frac{1}{{\partial}_-^{\;\;2}}{\Delta}(x;m^2)-
\frac{x^-}{m^2-n_-{\partial}_-^{\;\;2}}
{\delta}(x^+)+\frac{2n_+{\partial}_+}{(m^2-n_-{\partial}_-^{\;\;2})^2}
{\delta}(x^+).
\end{equation}
It remains to be shown whether we can use (\ref{eq:(4.1)}) to obtain 
consistent pure space-like axial gauge quantization conditions which are 
needed to quantize interacting pure space-like axial gauge fields. We 
leave these tasks for subsequent studies.

\appendix
\section{derivation of {\rm $(3{\cdot}16)$}}
In this appendix we give a detailed derivation of the $x^+$-ordered gauge 
field propagator
\begin{equation}
D_{{\mu}{\nu}}(q)=\int d^2x{\langle}0|\{{\theta}(x^+)A_{\mu}(x)
A_{\nu}(0)+{\theta}(-x^+)A_{\nu}(0)A_{\mu}(x) \}|0{\rangle}e^{iq{\cdot}x}.
\label{eq:(A.1)} 
\end{equation}
It is straightforward to show that the contributions from the 
$\tilde{\Sigma}$ and from the residual gauge fields are given, 
respectively, by
\begin{equation}
D^p_{++}(q)=\frac{i}{q^2-m^2+i{\varepsilon}}\frac{m^2}{q_-^{\;\;2}}
=\frac{i}{q^2-m^2+i{\varepsilon}} \left( -g_{++}
+\frac{2n_+q_+}{q_-}-n^2\frac{q_+^{\;\;2}}{q_-^{\;\;2}} 
\right)-\frac{i}{q_-^{\;\;2}},  \label{eq:(A.2)} 
\end{equation}
\begin{eqnarray}
&D^g_{++}(q)={\int}^{\infty}_0dk_+\Bigl[ \Bigr. 
{\delta}^{\prime}(q_-)\frac{n_-k_+^{\;\;2}}{m^2+n_-k_+^{\;\;2}}
\Bigl(\frac{i}{k_+-q_+-i{\varepsilon}}-\frac{i}{k_++q_+-i{\varepsilon}}\Bigr) &
\nonumber \\
&+ {\delta}(q_-) \frac{2n_+k_+m^2}{(n^2k_+^2+m^2)^2}
\left( \frac{i}{k_+-q_+-i{\varepsilon}}+\frac{i}{k_++q_+-i{\varepsilon}}\right)
\Bigl.\Bigr], \label{eq:(A.3)}
\end{eqnarray}
Here, we have made use of the fact that $q^2=n_-q_-^{\;\;2}+2n_+q_+q_-
-n_-q_+^{\;\;2}$ and that $n^2=n_-$.
Note that the explicit $x^-$ dependence gives rise to the factor ${\delta}^
{\prime}(q_-)$. Note also that there is no on-mass-shell condition for 
the residual gauge fields so that there remains a $k_+$-
integration. As a consequence, there arise singularities resulting from the 
inverse of the hyperbolic Laplace operator. It turns out that when we 
regularize the singularities by PV regularization, the integral on the first 
line of (\ref{eq:(A.3)}) is well-defined. In fact we can rewrite its integrand 
as a sum of simple poles:
\begin{eqnarray}
&\frac{n_-k_+^{\;\;2}}{m^2+n_-k_+^{\;\;2}}\left(\frac{i}
{k_+-q_+-i{\varepsilon}}-\frac{i}{k_++q_+-i{\varepsilon}}\right)
=\frac{n_-q_+^{\;\;2}}{m^2+n_-q_+^{\;\;2}}\left(\frac{i}{k_+-q_+-i
{\varepsilon}} \right.& \nonumber \\
&-\left.\frac{i}{k_++q_+-i{\epsilon}}\right) -\frac{n_-a}
{m^2+n_-q_+^{\;\;2}}\left(\frac{i}{k_+-a}-\frac{i}{k_++a}\right),&
\label{eq:(A.4)}
\end{eqnarray}
where $a=\frac{m}{\sqrt{-n_-}}$. Direct calculation then gives 
\begin{equation}
{\int}^{\infty}_0dk_+\left( \frac{i}{k_+-q_+-i{\varepsilon}}
-\frac{i}{k_++q_+-i{\varepsilon}} \right)=-{\pi}{\rm sgn}(q_+), 
\label{eq:(A.5)}
\end{equation}
\begin{equation}
{\rm P}{\int}^{\infty}_0dk_+\left( \frac{1}{k_+-a}-\frac{1}{k_++a} \right)=0,
\label{eq:(A.6)}
\end{equation}
where ${\rm sgn}(q_+)$ is obtained because the $k_+$-
integration region is limited to be $(0,{\infty})$. On the other hand, the integral on 
the second line of (\ref{eq:(A.3)}) yields a linear divergence. We can see that 
by rewriting the integrand as a sum of simple and double poles:
\begin{eqnarray}
&{}& \frac{2n_+k_+m^2}{(m^2+n_-k_+^{\;\;2})^2}\left(\frac{i}{k_+-q_+-
i{\varepsilon}}+\frac{i}{k_++q_+-i{\varepsilon}}\right) 
=\frac{2n_+q_+m^2}{(m^2+n_-q_+^{\;\;2})^2} \nonumber \\
&{\times}&\left(\frac{i}{k_+-q_+-i{\varepsilon}}
-\frac{i}{k_++q_+-i{\varepsilon}}\right) 
+n_+a\frac{n_-q_+^{\;\;2}-m^2}{(m^2+n_-q_+^{\;\;2})^2} 
\left(\frac{i}{k_+-a}-\frac{i}{k_++a}\right) \nonumber \\
&-&\frac{n_+}{n_-}\frac{m^2}{m^2+n_-q_+^{\;\;2}}\left( \frac{i}{(k_+-a)^2}
+\frac{i}{(k_++a)^2}\right).
\label{eq:(A.7)}
\end{eqnarray}
We can evaluate the integrations of the first and second terms on the right hand with the help of (\ref{eq:(A.5)}) and (\ref{eq:(A.6)}). However, we 
cannot regularize a linear divergence resulting from the double pole by the PV 
prescription. We show below that this linear divergence cancels one resulting 
from the factor $(q_-^{\;\;2})^{-1}$ of $D_{++}^p$ in (\ref{eq:(A.2)}). For 
later convenience we rewrite the linearly diverging integration in the form
\begin{equation}
{\rm P}{\int}^{\infty}_0dk_+ \left( \frac{n_+}{(k_+-a)^2}+\frac{n_+}
{(k_++a)^2} \right)={\rm P}{\int}^{\infty}_{-\infty}dk_+\frac{n_+}{(k_+-a)^2}
=-\int^{\infty}_{-\infty}dq_-\frac{n_-}{q_-^{\;\;2}},
\label{eq:(A.8)}
\end{equation}
where we have  changed the integration variable from $k_+$ to 
$q_-=\frac{-n^2}{n_+}(k_+-a)$. Substituting  (\ref{eq:(A.5)}), 
(\ref{eq:(A.6)}) and  (\ref{eq:(A.8)}) into  (\ref{eq:(A.3)}) yields
\begin{eqnarray}
&D_{++}^g(q)=\frac{i}{q^2-m^2+i{\epsilon}}\Bigl( \Bigr.
-in^2q_+^{\;\;2}{\pi}{\rm sgn}(q_+){\delta}^{\prime}(q_-)
-2in_+q_+{\delta}(q_-){\pi}{\rm sgn}(q_+) &  \nonumber \\
&\left.-m^2{\delta}(q_-){\int}^{\infty}_{-\infty}dq_-\frac{1}
{q_-^{\;\;2}} 
\right),&  \label{eq:(A.9)}
\end{eqnarray}
where we have made use of the identity
\begin{equation}
\frac{1}{q^2-m^2+i{\varepsilon}}{\delta}^{\prime}(q_-)=
-\frac{1}{m^2+n_-q_+^{\;\;2}}{\delta}^{\prime}(q_-)
+\frac{2n_+q_+}{(m^2+n_-q_+^{\;\;2})^2}{\delta}(q_-). \label{eq:(A.10)}
\end{equation}
Thus, for the sum of (\ref{eq:(A.2)}) and (\ref{eq:(A.9)}) we obtain
\begin{eqnarray}
&D_{++}(q)=\frac{i}{q^2-m^2+i{\epsilon}}\Bigl( \Bigr.
-g_{++}+\frac{2n_+q_+}{q_-+i{\varepsilon}{\rm sgn}(q_+)} -in^2q_+^{\;\;2}{\pi}
{\rm sgn}(q_+){\delta}^{\prime}(q_-) &  \nonumber \\
&\left.
-n^2q_+^{\;\;2}(\frac{1}{q_-^{\;\;2}}-{\delta}(q_-){\int}^{\infty}_{-\infty}
dq_-\frac{1}{q_-^{\;\;2}} \right) -i(\frac{1}{q_-^{\;\;2}}-{\delta}(q_-){\int}^
{\infty}_{-\infty}dq_-\frac{1}{q_-^{\;\;2}}),&  \label{eq:(A.11)}
\end{eqnarray}
where we have made use of the identity:
\begin{equation}
{\delta}(q_-)\frac{-m^2}{q^2-m^2+i{\varepsilon}}={\delta}(q_-)\left(
1+\frac{n^2q_+^{\;\;2}}{q^2-m^2+i{\varepsilon}} \right). \label{eq:(A.12)}
\end{equation}

Now we show that the term $\frac{1}{q_-^{\;\;2}}-{\delta}(q_-){\int}^{\infty}_
{-\infty}dq_-\frac{1}{q_-^{\;\;2}}$ does not give rise to any divergences 
when we calculate $D_{++}(x)$ by substituting  (\ref{eq:(A.11)}) into 
(\ref{eq:(A.1)}). To show this we consider the integration: 
\begin{eqnarray}
&{\int}^{\infty}_{-\infty}dq_-\frac{1}{q^2-m^2+i{\varepsilon}}\left( 
\frac{1}{q_-^2}
-{\delta}(q_-){\int}^{\infty}_{-\infty}dq_-\frac{1}{q_-^2}\right)e^{-iq_-x^-}& 
\nonumber \\
&={\int}^{\infty}_{-\infty}dq_-\frac{e^{-iq_-x^-}}{q_-^{\;\;2}(q^2-m^2+
i{\varepsilon})}+
\frac{1}{m^2+n_-q_+^{\;\;2}}{\int}^{\infty}_{-\infty}dq_-\frac{1}{q_-^2}.&  
\label{eq:(A.13)}
\end{eqnarray}
We can rewrite the second line further as
\begin{eqnarray}
&{}&\!{\int}^{\infty}_{-\infty}dq_-\frac{e^{-iq_-x^-}-1}{q_-^{\;\;2}}\frac{1}
{q^2-m^2+i{\varepsilon}}+\hspace{-1.5mm}
{\int}^{\infty}_{-\infty}dq_-\frac{1}{q_-^{\;\;2}}
\left( \frac{1}{ m^2+n_-q_+^{\;\;2}}+\frac{1}{q^2-m^2+i{\varepsilon}}\right)
\nonumber \\
&=&\hspace{-2mm}
{\int}^{\infty}_{-\infty}dq_-\left(\frac{e^{-iq_-x^-}-1}{q_-^{\;\;2}}
\frac{1}{q^2-m^2+i{\varepsilon}}+\frac{2n_+q_++n_-q_-}{q_-(m^2+n_-q_+^{\;\;2})
(q^2-m^2+i{\varepsilon})}\right).  \label{eq:(A.14)}
\end{eqnarray}
We see that the last integrals diverge at most logarithmically; but 
logarithmic divergences can be regularized by the PV prescription so
there arise no divergences in (\ref{eq:(A.14)}). This verifies that 
the following identity holds:
\begin{eqnarray}
&\frac{1}{q_-^2}+i{\pi}{\rm sgn}(q_+){\delta}^{\prime}(q_-)
-{\delta}(q_-){\int}^{\infty}_{-\infty}dq_-\frac{1}{q_-^2}& \nonumber \\
&={\rm Pf}\frac{1}{q_-^2}+i{\pi}{\rm sgn}(q_+){\delta}^{\prime}(q_-)
=\frac{1}{(q_-+i{\epsilon}{\rm sgn}(q_+))^2}&  \label{eq:(A.15)}
\end{eqnarray}
where Pf denotes Hadamard's finite part. It should be also noted that
\begin{equation}
\frac{1}{2{\pi}}{\int}^{\infty}_{-\infty}dq_-\left({\delta}(q_-)
{\int}^{\infty}_{-\infty}dq_-\frac{1}{q_-^2}-\frac{1}{q_-^2}\right)
e^{-q_-x^-}=\frac{1}{{\pi}}{\int}^{\infty}_0dq_-\frac{1-{\rm cos}q_-x^-}{q_-^2}
=\frac{|x^-|}{2}, \label{eq:(A.16)}
\end{equation}
where the Fourier transform of the last term is $-\frac{1}{2}(\frac{1}{(q_-+i
{\epsilon})^2}+\frac{1}{(q_--i{\epsilon})^2})$. Thus we obtain
\begin{eqnarray}
&D_{++}(q)=\frac{i}{q^2-m^2+i{\varepsilon}}\left(-g_{++}+
\frac{2n_+q_+}{q_-+i{\epsilon}{\rm sgn}(q_+)}-\frac{n^2q_+^{\;\;2}}
{(q_-+i{\varepsilon}{\rm sgn}(q_+))^2} \right)& \nonumber \\ 
&-{\delta}_{{\mu}+}{\delta}_{{\nu}+}\frac{i}{2}
\left(\frac{1}{(q_-+i{\epsilon})^2}+\frac{1}{(q_--i{\epsilon})^2}\right).&
 \label{eq:(A.22)}
\end{eqnarray}
This completes showing that, due to the residual gauge fields, the linear 
divergence resulting from the ${\partial}_+^{-1}$ is eliminated from the $x^+$-
ordered gauge field propagator.

\end{document}